\documentstyle[a4,eqsecnum,aps,epsf]{revtex}
\def\rhocol{\rho_{\mathrm{col}}}
\def\Ncol{N_{\mathrm{col}}}
\def\kappacol{\kappa_{\mathrm{col}}}
\def\etacol{\eta_{\mathrm{col}}}
\def\moyenne#1{\left<#1\right>}
\def\fexec{f_{\mathrm{exec}}}
\def\Csel{C_{\mathrm{salt}}}
\def\lambdasel{\lambda_{\mathrm{salt}}}
\def\lambdacol{\lambda_{\mathrm{col}}^{\mathrm{eff}}}
\def\deltaz{\Delta z^*}

\begin{document}

\title{\def\thefootnote{}Equilibrium sedimentation profiles of charged
colloidal suspensions\footnote{LPENS-Th 10/99}}
\author{Gabriel T\'ellez and Thierry Biben} \address{Laboratoire de
Physique (Unit\'e Mixte de Recherche du Centre National de la
Recherche Scientifique - UMR 5672) Ecole Normale Sup\'erieure de Lyon,
69364 Lyon cedex 07, France}
%\date{\today}
\maketitle

\begin{abstract}
We investigate the sedimentation equilibrium of a charge stabilized
colloidal suspension in the regime of low ionic strength. We 
analyze the asymptotic behaviour of the density profiles on the basis
of a simple Poisson--Boltzmann theory and show that the effective
mass we can deduce from the barometric law corresponds to the actual
mass of the colloidal particles, contrary to previous studies.
\end{abstract} 


\section{Introduction}

Under the action of gravity a colloidal suspension sediments 
to form a stratified fluid. The equilibrium density profile
of the colloidal particles results from the balance between 
the gravitational force and thermodynamic forces as derived from
the free energy of the system. The density profiles usually
exhibits a dense layer of colloidal particles at the bottom of
the container above which a light cloud of colloidal particles 
floats. In this last regime, the density of particles is small
enough to treat the fluid as an ideal gas. Under the reasonable assumption
that density gradients can be neglected, the equilibrium 
colloidal density obey the well known barometric law:
\begin{equation}
\rhocol(z)=\rhocol^0 \exp(-z/l_g)
\end{equation}
Here, $\rhocol(z)$ denotes the density profile of the colloidal particles,
$z$ is the altitude and $l_g=(\beta M g)^{-1}$ is the gravitational length
where $\beta=(k_{\mathrm{B}} T)^{-1}$ is the inverse temperature,
$M$ is the buoyant mass of a colloidal particle and $g$ the intensity of
the gravitational field. This exponential law is of practical interest
since it gives a prescription for the measurement of the buoyant mass 
$M$ of the particles. However a recent experimental study of the
sedimentation profiles of strongly de-ionized charged colloidal suspensions
\cite{Piazza} lead the authors to challenge the validity of this barometric law.
An exponential behaviour was indeed observed in the asymptotic regime, but 
the measured gravitational length $l_g^*$ could differ significantly from
the expected one (a factor of two). $l_g^*$ was found to systematically 
overestimate the actual value $l_g$, with the result that the buoyant mass
measured within these experiments is systematically reduced compared to the
known buoyant mass of the particles.

Some theoretical efforts have been made to study this problem. First
Biben and Hansen~\cite{BibHans} solved numerically the problem in a
mean field approach, but unfortunately due to numerical difficulties
the samples height considered where of the order of the micron while
in the experiment the samples height are of the order of the
centimeter. As a consequence, the dilute region at high altitude could
 not be studied
in this approach. Nevertheless the numerical results show a positive
charge density at the bottom of the container and a negative charge at
the top while the bulk of the container is neutral. This result show
that a non-zero electric field exists in the bulk of the container and
acts against gravity for the colloids.

More recently one of the authors studied a two-dimensional solvable
model for this problem~\cite{tcp+g}. This model is not very realistic
(the valency of the colloids was $Z=1$ and there was no added salt)
but has the nice feature of being exactly solvable analytically. It
confirmed the condenser effect noticed for small height containers in
Ref.~\cite{BibHans}. For large height containers it showed a new
interesting phenomenon: while there is still a positive charge density
at the bottom of the container, the negative charge density is not any
more at the top of the container floating but at some altitude.
Interestingly, the analytical expression for the density profiles in
the asymptotic regime predicts a decay in $\exp(-z/l_g)/z$ for the
colloidal density. Besides the $1/z$ factor that cannot be explained
by a mean field approach, no mass reduction is predicted by this
model. However one should be cautious when comparing two-dimensional
systems to the three dimensional case because the density in not
relevant in two-dimensional Coulomb systems: no matter how small the
density is the system is always coupled, the ideal gas regime is never
attained. For this reason a decay of the density similar to the one of
an ideal gas is in itself surprising in two dimensions.

Lately new results based on an approximate version of the model introduced
in reference \cite{BibHans} lead the authors of these studies 
\cite{Simonin,Lowen} to conclude that the mean-field approach was indeed
able to predict a mass reduction in the asymptotic regime. 
Here we present some new results about this problem treated under the
Poisson-Boltzmann approximation, and show that it is indeed not the
case.

\section{The model and the Poisson-Boltzmann approximation}

Let us consider some
colloidal particles (for example some latex spheres) in a solution
with some amount of added salt. In a polar solvent like water the
colloids release some counterions and therefore acquire a surface
electric charge $Ze$ ($Z$ is a entire number usually positive and $-e$
is the charge of the electron). We consider that the colloidal sample
is monodisperse, all colloids have the same valency $Z$, and that
the counterions and the salt cations are both monovalent and therefore
we shall not make any distinction between cations coming from the
colloids and salt cations. 
We then consider a three-component system composed of colloidal particles
with electric charge $Ze$ and mass $M$, counterions with charge $-e$
and coions with charge $+e$. We shall neglect the masses of the
counterions and coions when compared with the mass of the colloids. The
solvent shall be considered in a primitive model representation as a
continuous medium of relative dielectric permittivity $\epsilon$ (for
water at room temperature $\epsilon\approx 80$). The system is in a
container of height $h$, the bottom of the container is at $z=0$
altitude. We consider that the system is invariant in the horizontal
directions. The density profiles of each species are denoted by
$\rhocol(z)$, $\rho_+(z)$ and $\rho_-(z)$ ($z$ is the vertical coordinate) 
for the colloids, the cations and the anions respectively at equilibrium.
Let us define the electric charge density (in units of
$e$) $\rho=Z\rhocol-\rho_-+\rho_+$ and the electric potential $\Phi$,
solution of the Poisson equation
\begin{equation}\label{Poisson}
\frac{d^2\Phi}{dz^2}(z)=-\frac{4\pi}{\epsilon} e\rho(z)
\end{equation}

It is instructive to recall that the Poisson-Boltzmann equation can be
derived from the minimization of the free energy density functional
\begin{eqnarray}\label{FreeEnergy}
{\cal{F}}[\rhocol,\rho_+,\rho_-]&=&
\sum_{i\in\{\mathrm{col},+,-\}}\int_0^h k_{\mathrm{B}}T \rho_i(z)
\left[\ln(\lambda_i^3\rho_i(z))-1\right]\,dz
\nonumber\\
&&+\int_0^h Mgz\rhocol(z)\,dz
+\frac{1}{2}\int_0^h e\rho(z)\Phi(z)
\end{eqnarray}
where $\lambda_i$ is the de Broglie wavelength of species $i$.
Minimization of the grand potential with respect to the densities: $\delta {\cal
F} / \delta \rho_i(z) - \mu_i = 0$, where $\mu_i$ is the chemical
potential of species $i$, yields
\begin{mathletters}
\label{Boltzmann}
\begin{eqnarray}
\rhocol(z)&=&\rhocol^0 \exp(-\beta Ze\Phi(z)-\beta M g z)\\
\rho_+(z)&=&\rho_+^0 \exp(-\beta e \Phi(z))\\
\rho_-(z)&=&\rho_-^0 \exp(\beta e \Phi(z))
\end{eqnarray}
\end{mathletters}
We shall work in the canonical ensemble, the prefactors $\rho_i^0$
which depend on the chemical potentials $\mu_i$ are determined by the
normalizing conditions
\begin{equation}
\int_0^h \rho_i(z)\, dz = N_i
\end{equation}
where $N_i$ is the total number of particles per unit area of
species $i$. The system is globally neutral so we have
$Z\Ncol-N_-+N_+=0$.

Let us introduce the following notations: $l_g=(\beta M g)^{-1}$ is
the gravitational length of the particles, $l=\beta e^2 / \epsilon$ is
the Bjerrum length, $\phi=\beta e \Phi$ is the dimensionless electric
potential and $\kappa_i=(4\pi l N_i/h)^{1/2}$. $\kappa_\pm^{-1}$ are the
Debye lengths associated to the counterions and the coions and
$(Z\kappacol)^{-1}$ is the Debye length associated to the colloidal
particles. For a quantity $q(z)$ depending on the altitude, let us
define its mean value $\moyenne{q}=\int_0^h q(z)\,dz/h$. With these
notations equations~(\ref{Poisson}) and~(\ref{Boltzmann}) yield the
modified Poisson-Boltzmann equation
\begin{equation}\label{Poisson-Boltzmann}
\frac{d^2\phi}{dz^2}(z)=-Z\kappacol^2 
\frac{e^{-Z\phi(z)-z/l_g}}{\moyenne{e^{-Z\phi(z')-z'/l_g}}}
+\kappa_-^2\frac{e^{\phi(z)}}{\moyenne{e^{\phi(z')}}}
-\kappa_+^2\frac{e^{-\phi(z)}}{\moyenne{e^{-\phi(z')}}}
\end{equation}
From Eq.~(\ref{Poisson-Boltzmann}) it is clear that the problem has
the following scale invariance: if $\phi(z)$ is a solution
of~(\ref{Poisson-Boltzmann}) then $\phi(\alpha z)$ is a solution of
the problem with the rescaled lengths $\alpha l_g$ and
$\alpha\kappa_i^{-1}$. 

The advantage of the density functional formulation of the problem is
that it allows for systematic corrections to the Poisson-Boltzmann
approximation. For instance, one may be interested in the effect of
the finite size of the macroions. Let $\sigma$ be the diameter of the
colloids, $\etacol=\pi\sigma^3\rhocol/6$ the volume fraction of
colloids and $\rho_\pm^*=\rho_\pm/(1-\etacol)$ the effective densities
of the microions. Then, the finite size of the colloids can be
accounted in a local density approximation (LDA) by adding to the free
energy density functional~(\ref{FreeEnergy}) the free energy excess
term
\begin{equation}
\int_0^h \fexec(\rhocol(z))\, dz
\end{equation}
where $\fexec(\rhocol)$ is the excess free energy of a hard sphere fluid
derived by the Carnahan--Starling equation of
state~\cite{CarnahanStarling}
\begin{equation}
\fexec(\rhocol)=k_{\mathrm{B}}T \rhocol
\frac{4\etacol-3\etacol^2}{(1-\etacol)^2} 
\end{equation}
and by replacing in~(\ref{FreeEnergy}) the entropy term of the
microions by
\begin{equation}
k_{\mathrm{B}}T\int_0^h
\rho_\pm(z)(\ln(\lambda_\pm^3\rho_\pm^*(z))-1)\, dz
\,.
\end{equation}
Minimization of this new free energy functional gives the
modified version of Eqs.~(\ref{Boltzmann})
\begin{mathletters}
\label{Boltzmann-mod}
\begin{eqnarray}
\rhocol(z)&=&\rhocol^0 \exp(
-\beta Ze\Phi(z)-\beta M g z)
\nonumber\\
&&\times\exp\left[
\frac{3\etacol(z)^3-9\etacol(z)^2+8\etacol(z)}{(1-\etacol(z))^3}
+\frac{\pi\sigma^3}{6}\left(\rho_+^*(z)+\rho_-^*(z)\right)
\right]
\\
\rho_+(z)&=&\rho_+^0 (1-\etacol(z))\exp(-\beta e \Phi(z))\\
\rho_-(z)&=&\rho_-^0 (1-\etacol(z))\exp(\beta e \Phi(z))
\end{eqnarray}
\end{mathletters}

There are several length scales in this problem: the gravitational
length of the colloids $l_g$, the Debye or screening length, the
height $h$ of the container and eventually the hard core diameter of
the particules. In a realistic case $h$ is of the order of the
centimeter, $l_g$ of the order of $0.1$ mm, the screening length of
the order of 10 nm. We are faced to a practical numerical problem,
when we will transpose the problem to a lattice, the lattice spacing
should be smaller than all the physical lengths, but since $h$ is much
larger than the others lengths, the number of sites in the lattice
should be very high (of order $10^6$). A possible approach to deal
with this problem is to study small containers as in
Ref.~\cite{BibHans}. In this paper we want to study the case of high
containers so we will consider very deionized systems in which the
screening length is of the order of 0.1 mm, much larger than in usual
physical cases, the other lengths taking ``physical'' values. That way
the necessary number of points in the lattice will be reasonable (a
few hundreds). Also, since the screening length is so large, the hard
core of the macroions will not change the results drastically from the
case of point particles so we will concentrate from now on on the
Poisson-Boltzmann problem for point
particles~(Eq.~\ref{Poisson-Boltzmann}).

\section{Results}

Equation~(\ref{Poisson-Boltzmann}) is solved numerically by an
iterative method~\cite{Badiali}. Using the Green function of the
one-dimensional Laplacian
\begin{equation}
G(z,z')=\frac{1}{2}\left|z-z'\right|
\end{equation}
the Poisson equation~(\ref{Poisson}) can be written as
\begin{equation}\label{Poisson-integrale}
\phi(z)=-4\pi l\int_0^h G(z,z')\rho(z')\,dz' 
\end{equation}
Starting with an arbitrary electric potential, one can compute the
corresponding density profiles using Eqs.~(\ref{Boltzmann}) and derive
a new electric potential using Eq.~(\ref{Poisson-integrale}), then
reiterate the process until a stationary solution is attained. In
practice instead of using the new potential directly for the next
iteration a mixing of the old and new densities is used.

\subsection{Generic results}

\begin{figure}
\epsfxsize=100mm
\hskip 30mm\epsffile{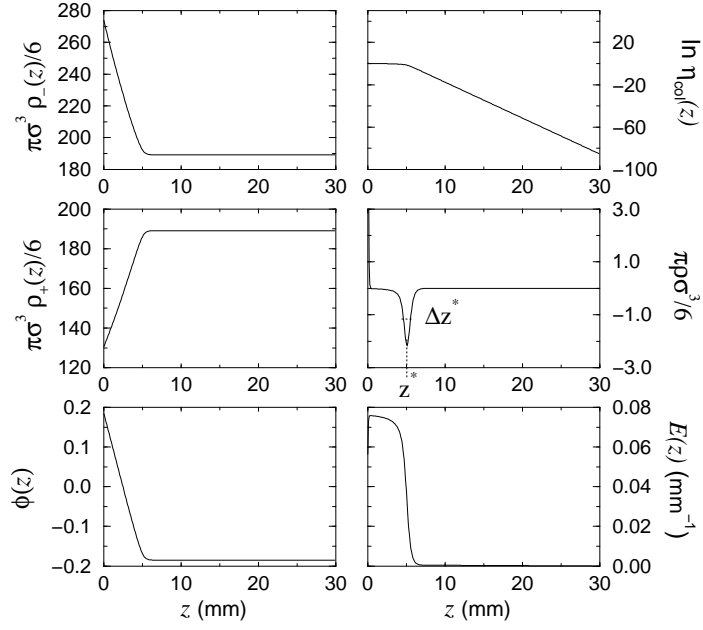}
\caption{\label{fig:tous}
From left to right, up to down, starting at the upper left corner,
profiles of: counterions density, colloidal volume fraction in natural
logarithm scale, coions density, charge density, dimensionless
electric potential $\phi$, electric field $E=-d\phi/dz$.  The
parameters used are: $l=7\cdot10^{-8}$~\AA, $l_g=0.128$~mm, $h=30$~mm,
$Z=100$, $\Csel=0.1$~mMol/l, $\bar\etacol=0.12$ and $\sigma=180$~nm.  
}
\end{figure}

As stated before we have to consider very deionized systems in which
the Bjerrum length $l$ is smaller than $10^{-6}$ \AA\ (the physical
value of $l$ for water at room temperature is 7
\AA). Figure~\ref{fig:tous} shows the density profiles of each species, the
charge density, the electric potential and the electric field profile
of a typical sample with the following parameters:
$l=7\cdot10^{-8}$~\AA, $l_g=0.128$ mm, $h=30$ mm, $Z=100$, a salt
concentration $\Csel=0.1$ mMol/l and a mean colloidal volume fraction
$\bar\etacol=0.12$ (we consider that the particles have a hard core
diameter $\sigma=180$~nm to express the colloidal density as a volume
fraction in order to use units familiar with the experiments but we do
not account for hard core effects in the Poisson--Boltzmann equation).

The log plot of colloidal density profiles is similar to the
experimental ones~\cite{Piazza}. In the bottom there is a slow decay
whereas at high altitudes there is a faster barometric decay. Since we
did not take into account the hard core of the particles in the theory
we do not find the discontinuity in the density profiles near the
bottom of the sample observed in the experiments~\cite{Piazza}, due to
the phase transition of the colloids from an amorph solid to a
fluid. At very low altitudes (near the bottom, very high volume
fractions) the Poisson--Boltzmann theory is not valid.

The charge density profile confirms the results of Ref.~\cite{tcp+g},
that there is a strong accumulation of positive charges at the bottom
of the container while there is a cloud of negative charge density
floating at some altitude $z^*$. There are clearly two neutral regions
in the container: one at low altitude between the positive charge
density at the bottom and the negative cloud, in which a non-vanishing
electric field exists and a second neutral region at high altitude,
over the negative cloud. The electric field in the lower region acts
against gravity for the colloids therefore, as seen in the log plot of
the colloids density profile, the decay is much slower than the one
for an ideal neutral gas. Numeric results for other series of samples
suggest that this electric field is proportional to $Mg/Z$.  In the
upper region the colloidal density drops exponentially as
$\exp(-z/l_g)$ since the electric potential is almost constant and the
electric field vanishes.

Since the different densities vary with the altitude we can define a
local screening length which depends on the altitude by
\begin{equation}
\lambda(z)=\left(
4\pi\l\left(Z^2 \rhocol(z)+\rho_+(z)+\rho_-(z)\right)
\right)^{-1/2}
\end{equation}
The two regions of the sediment are caracterized by a very different
behavior of this local screening length. In the lower region the
colloidal density is so high that $Z^2 \rhocol(z) \gg
\rho_+(z)+\rho_-(z)$. In that region the colloids dominate the
screening length. On the other hand in the upper region the colloidal
density is very small and salt now controles the screening
length which is then constant since at high altitudes the cations and
anions densities are almost constant and equal to the mean salt
concentration as seen in figure~\ref{fig:tous}.  It is interesting to
notice that electric charges accumulate in the intermediate region
around $z^*$ where there is a change of regime, in agreement with
macroscopic electrostatics principles.

The preceding remark allows us to understand how the physical
parameters (mean volume fraction of colloids, mass of the colloids,
amount of added salt) will modify the altitude $z^*$ which separates
the two regions. For example if we add more salt, $z^*$ will diminish
since we reach sooner the regime where $Z^2 \rhocol(z) <
\rho_+(z)+\rho_-(z)$. We have computed the density profiles in several
other cases changing the values of the parameters in order to find the
depency of $z^*$ in these parameters. Our numerical results suggest
that 
\begin{equation}
z^*=-\frac{c_1}{\sqrt{l\Csel}} + a_2 \frac{Z \sqrt{\Ncol l_g}}{\sqrt{\Csel}}
\end{equation}
with $c_1=0.15\pm0.05$ and $a_2=1.0\pm0.1$.
The preceding equation can be written in a more attracting way,
introducing the screening length associated to the salt
$\lambdasel=(4\pi l \Csel)^{-1/2}$ and the effective screening length
associated to the colloids $\lambdacol=(4\pi l Z^2 \Ncol/l_g)^{-1/2}$,
as
\begin{equation}
z^*=\lambdasel\left( -a_1 + a_2 \frac{l_g}{\lambdacol} \right)
\end{equation}
with $a_1=\sqrt{4\pi}c_1=0.5\pm0.2$. We do not
consider here boundary effects: this equation is only valid if $z^*$
is smaller than $h$. The finite height $h$ of the container will have
the effect to ``push'' the negative cloud downwards if the parameters
are such that $z^*$ approaches the top of the container. The same holds
for the bottom of the container if $z^*$ is too small.

Another quantity of interest is the size $\deltaz$ of the negative
cloud, defined as the mid-height width of the negative peak in the
charge density profile (see figure~\ref{fig:tous}). Since we know that
at $z^*$ altitude, $Z^2 \rhocol(z^*)$ is of the same order of
magnitude as $\rho_+(z^*)+\rho_-(z^*)=2\Csel$, the screening length at
that altitude is proportional to $\lambdasel$. From basic
electrostatics we know that the system will only tolerate charges over
a length of order of the screening length, we deduce that $\deltaz$ is
proportional to $\lambdasel$. In fact the numerical results suggest
also a linear dependency of $\deltaz$ in $l_g$:
\begin{equation}
\deltaz=b_1 l_g + b_2 \lambdasel
\end{equation}
with $b_1=5.0\pm0.5$, $b_2=0.7\pm0.2$, and the same
restrictions concerning boundary effects as for the equation for
$z^*$.

\subsection{The apparent mass}

As we mentioned before, at high altitudes (larger than $z^*$) the
electric potential is almost constant and the electric field
vanishes. From this it is clear that the colloidal density will decay
as $\exp(-z/l_g)$ and there is no apparent reduced mass. Nevertheless
let us notice that in the regime where the electric potential is
almost constant in our calculations the corresponding colloidal volume
fraction is smaller than $10^{-9}$. Such volume fractions cannot be
measured experimentally. In practice the optical methods used
in~\cite{Piazza} allow to measure only volume fractions larger than
$10^{-5}$. A possible explanation to the apparent mass observed in the
experiments is that for volume fractions higher than $10^{-5}$ the
asymptotic regime where the electric field vanish have not been
reached yet: there is a residual electric field responsible of the
observed reduced mass.

In order to test this hypothesis we made a log plot of several
colloidal volume fraction profiles restricting the plot to volume
fractions higher than $10^{-5}$ (Figure~\ref{fig:pentes}).  We
computed the slope of the wing of the colloidal density to find an
effective gravitational length $l_g^*$ which is higher than the actual
gravitational length $l_g$ as observed in the experiments. Futhermore
when we plot the colloidal volume fraction profile and the
corresponding electric field profile together (Figure~\ref{fig:logeta-et-E})
we notice that for volume fractions higher than $10^{-5}$ the
electric field is not zero.

\begin{figure}
\epsfxsize=100mm
\hskip 30mm\epsffile{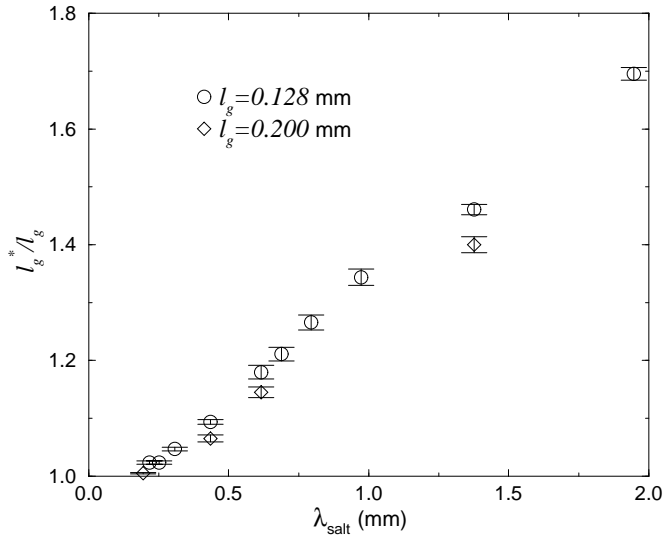}
\caption{\label{fig:lgstar-lambdasel} The ratio of apparent
gravitational length by the actual gravitational length $l_g^*/l_g$
versus the salt screening length $\lambdasel$, for two different
values of the gravitational length.}
\end{figure}

The different plots in Figure~\ref{fig:pentes} where obtained using different salt
concentrations, so the sediment height (which is proportional to
$z^*$) varies. In this case we found that the apparent mass is a
decreasing function of the height of the sediment, in agreement with
the experiments. However the sediment height can be changed by
changing other parameters like the mean colloidal density or their
valency $Z$. Computing the apparent gravitational length $l_g^*$ as
defined before for other series of samples, we found that the apparent
gravitational length $l_g^*$ does not depend on $Z$ or the mean
colloidal density. In our model the ratio $l_g^*/l_g$ is only a
function of the salt density. Figure~(\ref{fig:lgstar-lambdasel})
shows the ratio $l_g^*/l_g$ as a function of salt screening length
$\lambdasel$.

\section{Comparison with previous approaches}

As mentionned in the introduction the model presented above has
motivated several studies both numerically \cite{BibHans} and analytically
\cite{Simonin,Lowen}. The purpose of this section is to compare our numerical
results with the most achieved version of the theory presented in reference
\cite{Lowen}. This theoretical approach is based on a constrained
minimization of the free energy functional (\ref{FreeEnergy}) assuming
an exponential {\em ansatz\/} for the density profiles:
\begin{mathletters}
\label{Hartmut}
\begin{eqnarray}
\rhocol(z)&=&{\Ncol~a\over l_g} \exp(-a z/l_g)\\
\rho_+(z)&=&\Csel\\
\rho_-(z)&=&\Csel + {Z\Ncol~b\over l_g}\exp(-b z/l_g)
\end{eqnarray}
\end{mathletters}
With this parametrization $a=M^*/M$ is the ratio of the reduced mass
$M^*$ by the buoyant mass $M$ of a colloidal particle, $\Csel$ denotes
the fixed salt concentration, and $\Ncol$ is the fixed overall
colloidal density per unit area, i.e.  $\int_0^{+\infty} dz
\rhocol(z) \equiv \Ncol$. The system considered in
\cite{Lowen} is semi-infinite, $z=0$ corresponds to the bottom of the sample
and $h=+\infty$.
$a$ and $b$ are the two variational dimensionless parameters of the
theory, and the equilibrium density profiles $\rhocol(z)$ and $\rho_-(z)$ 
correspond to the values of $a$ and $b$ that minimize the free
energy functional (\ref{FreeEnergy}). Following reference \cite{Lowen}, the
minimization conditions express:
\begin{mathletters}
\label{Hartmut_min}
\begin{eqnarray}
b(a) = a\Biggl({2\over\sqrt{1+(1-a)/\gamma}} -1\Biggr)&&\\
Zb(a)-\kappa I\Biggl({Zb(a)\over \kappa}\Biggr) -\gamma +
{4\gamma b^2(a)\over (a+b(a))^2}&=&0
\end{eqnarray}
\end{mathletters}
where $\gamma = \pi Z^2 l\Ncol l_g$ is the coupling parameter
($l$ is the Bjerrum length introduced previously), and $\kappa = \Csel l_g/\Ncol$ 
is the relative amount of added salt. Function $I$ is defined by
$I(x) = \int_0^x dy(\ln(1+y))/y$. 

Although equations (\ref{Hartmut_min}) require a numerical treatment, it is
possible to extract asymptotic expressions when the coupling parameter
$\gamma$ is vanishingly small (strong gravitational coupling regime) or
large compared to unity (strong Coulomb coupling regime). Such an analysis
is presented in reference \cite{Lowen}, and we only reproduce here
the main features.
When gravitational coupling is strong, $\gamma\ll 1$, the reduced
mass is given by $a\simeq 1- 3\gamma$ (forall values of the salinity $\kappa$)
and therefore no mass reduction is observed in this regime (in agreement with
the numerical results presented in reference \cite{BibHans}). On the
contrary, in strong Coulomb coupling regimes, $\gamma\gg 1$, quite a large
mass reduction is predicted, even in low salinity regimes $\kappa\ll 1$
(in such a situation the mass reduction is given by $a\simeq\Bigl[1+{\kappa\over 2}
\ln^2(\kappa(1+{1\over Z}))\Bigr]/(Z+1) + O(1/\gamma)$).
Our numerical results based on a free minimization of the functional
(\ref{FreeEnergy}) show that it is indeed not the case, even though we observe
nice exponential asymptotic behaviours at high altitudes. To emphasize this
point we present in table \ref{table:res} data obtained mostly in the strong Coulomb 
coupling regime. Excepted for the first value $\gamma=7.2~10^{-3}$, which corresponds 
to the opposite strong gravitational coupling regime, the agreement between the
theory presented in reference \cite{Lowen} and the present free minimization is
very poor. The effective mass predicted by the free minimization procedure
corresponds to the actual mass within less than 0.1\% in most situations, whereas
the theory predicts effective masses that are only 2\% of the actual mass!

To our opinion, the failure of the parametrization presented in reference
\cite{Lowen} is not due to the exponential {\em ansatz\/} itself, but to the
constraint of global charge neutrality applied to the asymptotic regime.
The theory presented in reference \cite{Lowen} assumes that the profiles
are exponential from the bottom of the sample to the top, as a result the
free energy functional (\ref{FreeEnergy}) has to be minimized with the
constraint of global charge neutrality. However, the actual situation
is quite different. If we refer to the experimental work done by Piazza et al.
\cite{Piazza}, the exponential regime is reached only above a macroscopic layer
of strongly interacting colloidal particles. Data presented in the 
previous section (see e.g. figure~\ref{fig:tous}) resulting from a free
minimization of the functional also exhibit a dense macroscopic layer of colloidal
particles in the bottom of the cell, and these profiles cannot be simply
represented by a single exponential. This feature can be incorporated to
the model suggested by L\"owen, by splitting the cell into two parts.
The upper part of the cell (above a given altitude ``$z_o$'') corresponds to
the asymptotic region where the profiles can be accurately represented 
by an exponential, whereas below $z_o$ the profiles are more complicated.
As we can see, $z_o$ is defined by the condition that the profiles are
exponential above it. There is then no upper bound on the value of $z_o$
and the asymptotic profiles should not depend on its precise value.
As a result $z_o$ can be chosen arbitrarily large.
As a consequence, the part of the fluid located below $z_o$ can be
considered as a reservoir fixing the chemical potential of the ionic
species $\mu_{\mathrm{col}}$ and $\mu_-$ ($\mu_+$ is irrelevant since
the local density $\rho_+(z)$ is held fixed, and is thus not a variational 
parameter). Although the full system must be charge neutral, the 
asymptotic part above $z_o$ has no reason to be neutral. We are then
lead to minimize the free energy of the upper part of the cell in the
grand-canonical ensemble. Assuming that parametrization
(\ref{Hartmut}) is valid above $z_o$ the minimization equation associated to
the colloidal particles reads:
\begin{equation}
{\partial {\cal F}[\rhocol,\rho_+,\rho_-]\over\partial a} = 
\mu_{\mathrm{col}}{\partial N_{\mathrm{col}}\over\partial a}
\end{equation}
where ${\cal F}[\rhocol,\rho_+,\rho_-]$ is now the free energy functional
above $z_o$:
\begin{eqnarray}
{\cal{F}}[\rhocol,\rho_+,\rho_-]&=&
\sum_{i\in\{\mathrm{col},+,-\}}\int_{z_o}^{+\infty} k_{\mathrm{B}}T \rho_i(z)
\left[\ln(\lambda_i^3\rho_i(z))-1\right]\,dz
\nonumber\\
&&+\int_{z_o}^{+\infty} Mgz\rhocol(z)\,dz
+\frac{1}{2}\int_{z_o}^{+\infty} e\rho(z)\Phi(z)
\end{eqnarray}
and $N_{\mathrm{col}} = \int_{z_o}^{+\infty}\rhocol(z) dz$ is the number
of colloidal particles above $z_o$ per unit area.
After some algebra, this minimization equation can be written on the form:
%\begin{eqnarray}
%(a-1) &&\Biggl\{1+z_o^* +{z_o^*}^2 \Biggr\} 
%+ a\Biggl\{ {\mu_{\mathrm{col}}\over k_BT} -
%                  \ln\Bigl(\lambda_{\mathrm{col}} {\Ncol a\over l_g}\Bigr)
%         \Biggr\} z_o^*   
%\nonumber\\
%&& - \gamma\Biggl[e^{-z_o^*}(1- 2 z_o^*) - 4 a^2
%e^{- {z_o^*b/a}}\Biggl({1\over (a+b)^2}-
%{z_o^*\over b(a+b)}\Biggr) \Biggr] = 0
%\end{eqnarray}
\begin{eqnarray}
(a-1) &&\Biggl\{1+z_o^* +{z_o^*}^2 \Biggr\} 
+ a\Biggl\{ {\mu_{\mathrm{col}}\over k_BT} -
                  \ln\left(\lambda_{\mathrm{col}}^3 {\Ncol a\over l_g}\right)
         \Biggr\} z_o^*   
\nonumber\\
&& + \gamma\Biggl[e^{-z_o^*}(1+ 2 z_o^*) + 4 
e^{- {z_o^*b/a}}\Biggl({a^2\over (a+b)^2}
-\frac{1}{2}
-{z_o^* (a^2+b^2)\over b(a+b)}\Biggr) \Biggr] = 0
\end{eqnarray}
where $z_o^* \equiv a z_o/l_g$. We can easily check that when $z_o^*=0$
we recover the first equation of condition (\ref{Hartmut_min}).
As $z_o^*$ can be chosen arbitrarily large in our model, we easily see that this
equation implies $a=1$ (no mass reduction) and:
\begin{equation}
{\mu_{\mathrm{col}}\over k_BT} = 
\ln\Bigl(\lambda_{\mathrm{col}}{\Ncol \over l_g}\Bigr)
\end{equation}
This new version of the theory is consistent with our numerical results
predicting no mass reduction.

\section{Conclusion}

\begin{figure}
\epsfxsize=100mm
\hskip 30mm\epsffile{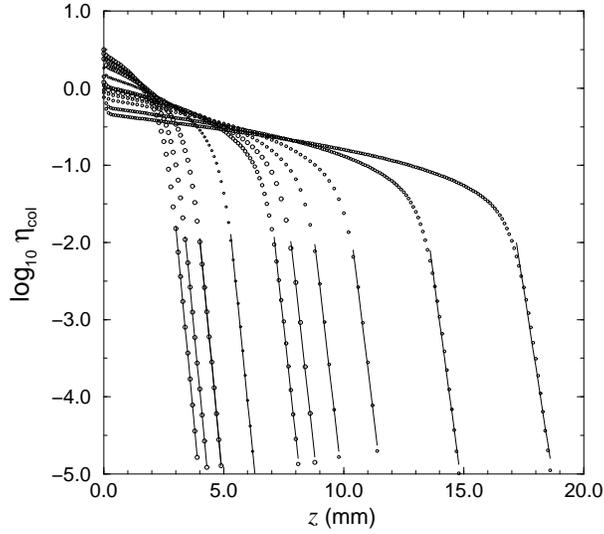}
\caption{\label{fig:pentes}
Colloidal density profile in decimal logarithm scale for different
salt concentrations and restricted to volume fractions higher than
$10^{-5}$. Common parameters to all curves are: $l_g=0.128$~mm,
$Z=100$, $\sigma=180$~nm, $\bar\etacol=0.12$ and $h=30$~mm.
The salt concentration in mMol/l from left to right is 0.4, 0.3, 0.2,
0.1, 0.05, 0.04, 0.03, 0.02, 0.01, 0.005. The apparent gravitational
length $l_g^*$ obtained from the slope of the low density wing is,
from left to right in mm: $0.131\pm0.003$, $0.131\pm0.003$, $0.134\pm0.003$,
$0.140\pm0.004$, $0.151\pm0.012$, $0.155\pm0.12$,
$0.162\pm0.013$, $0.172\pm0.014$, $0.187\pm0.009$, $0.217\pm0.011$.  
}
\end{figure}

\begin{figure}
\epsfxsize=100mm
\hskip 30mm\epsffile{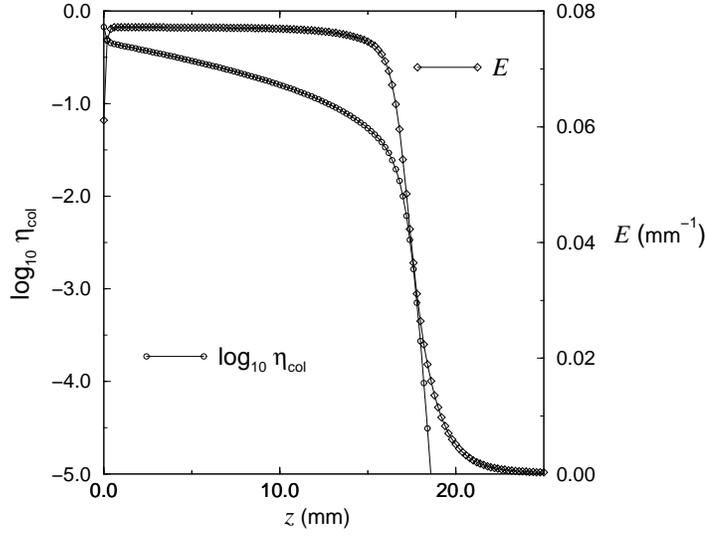}
\caption{\label{fig:logeta-et-E}
Colloidal volume fraction decimal logarithmic profile and the
corresponding electric field profile in the case $\Csel=0.005$~mMol/l,
the other parameters being those of figure~\ref{fig:pentes}. Notice that in
the low density wing used to compute the apparent gravitational
length~$l_g^*$ the electric field is not zero.
}
\end{figure}

A free minimization of the Poisson--Boltzmann theory used in references
\cite{BibHans,Lowen} have been performed in this article which lead us
to conclude that this simple mean field theory does not predict any
mass reduction contrarily to previous approximate minimization of the
same functional. These new results are fully consistent with the
analytical results obtained in a two-dimensional case by T\'ellez
\cite{tcp+g}. In particular, we observe the same condenser effect
between the bottom of the container and the top of the dense region,
resulting from a competition between electroneutrality and entropy of
the microions. Data plotted in figure~\ref{fig:pentes} give a possible
explanation for the experimental results obtained by Piazza et
al. \cite{Piazza}. Although in the asymptotic regime we observe no
mass reduction, this regime is attained for very low values of the
colloidal packing fractions, below the experimental resolution
($10^{-5}$) in some situations. As a result, the residual
electrostatic field can affect the profiles resulting in an apparent
effective mass.

\newpage

\newpage

\newpage

\begin{table}
\vspace{0.3cm}
{\centering \begin{tabular}{|c|c|r|c|c|c|}
\hline 
\( \gamma  \)&
\( \kappa  \)&
\( \Csel \)\( \; \; \; \; \; \;  \)&
\( Z \)&
\( M^*/M \) : Theory~\cite{Lowen} &
\( M^*/M \) : free minimization\\
\hline 
\( 7.2\: 10^{-3} \)&
\( 6\: 10^{4} \)&
\( 3.27\: 10^{-4} \) Mol / l&
\( 100 \)&
\( 0.9915 \)&
\( 0.9994\pm 0.0001 \)\\
\hline 
\( 33.2 \)&
\( 0.33 \)&
\( 5\: 10^{-6} \) Mol / l&
\( 100 \)&
\( 0.0204 \)&
\( 0.986\pm 0.015 \)\\
\hline 
\( 33.2 \)&
\( 6.5 \)&
\( 10^{-4} \) Mol / l&
\( 100 \)&
\( 0.0590 \)&
\( 0.999\pm 0.001 \)\\
\hline 
\( 33.2 \)&
\( 325 \)&
\( 5\: 10^{-3} \) Mol / l&
\( 100 \)&
\( 0.3103 \)&
\( 1.00001\pm 0.00001 \)\\
\hline 
\( 133 \)&
\( 6.5 \)&
\( 10^{-4} \) Mol / l&
\( 200 \)&
\( 0.0296 \)&
\( 0.9998\pm 0.0003 \)\\
\hline 
\end{tabular}\par}
\vspace{0.3cm}
\caption{\label{table:res}
Comparison between the reduced mass predicted by L\" owen's theory 
and our numerical results from the free minimization.
} 
\end{table}

%
% bio{#1=authors}{#2=Journal}{#3=vol. number}{#4=page}{#5=year}
%
\def\bio#1#2#3#4#5{{\rm #1, #2 }{\bf #3}{\rm, #4 (#5).}}


\begin{references}

\bibitem{Piazza} \bio{R.~Piazza, T.~Bellini, and V.~Degiorgio}%
{Phys.~Rev.~Lett.}{71}{4267}{1993}

\bibitem{BibHans} \bio{T.~Biben and J.~P.~Hansen}{J.~Phys.~C:
Condens.~Matter}{6}{A345}{1994}

\bibitem{tcp+g} \bio{G.~T\'ellez}{J.~Chem.~Phys.}{106}{8572}{1997} and
\bio{G.~T\'ellez}{J.~Phys.~A: Math.~Gen.}{31}{5277}{1998} 

\bibitem{Simonin} \bio{J.~P.~Simonin}{J.~Phys.~Chem.}{99}{1577}{1995}

\bibitem{Lowen} \bio{H.~L\"owen}{J.~Phys.:~Condens.~Matter}{10}{L479}{1998}


\bibitem{CarnahanStarling} \bio{N.~F.~Carnahan and
K.~E.~Starling}{J.~Chem.~Phys.}{51}{635}{1969} 

\bibitem{Badiali}\bio{J.~P.~Badiali, M.~L.~Rosinberg, D.~Levesque and
J.~J.~Weis}{J.~Phys.~C: Solid State Phys.}{16}{2183}{1983}



\end{references}
\end{document}